\documentclass{emulateapj}
\usepackage{apjfonts}

\newcommand{\msun}{M_\odot}

\newcommand{\mdot}{\dot{M}}
\newcommand{\ml}{M$_{\odot}$ yr$^{-1}$}

\newcommand{\gapprox}{\mathrel{\mathpalette\@versim>}}
\newcommand{\lapprox}{\mathrel{\mathpalette\@versim<}}
\newcommand{\propapprox}{\mathrel{\mathpalette\@versim\propto}}

\shorttitle{Mid-IR Non-Detection of SN 2010jl} 
\shortauthors{WILLIAMS \& FOX}

\begin{document}

\title{{\it SOFIA} Observations of SN 2010jl: Another Non-Detection of
  the 9.7 $\mu$m Silicate Dust Feature}

\author{Brian J. Williams,\altaffilmark{1}
Ori D. Fox,\altaffilmark{2}
}

\altaffiltext{1}{CRESST and X-ray Astrophysics Laboratory, NASA/GSFC, 8800 Greenbelt Road, Greenbelt, MD, Code 662, brian.j.williams@nasa.gov}
\altaffiltext{2}{Department of Astronomy, University of California, Berkeley, CA 94720-3411}

\begin{abstract}

We present photometric observations from the {\it Stratospheric
  Observatory for Infrared Astronomy (SOFIA)} at 11.1 $\mu$m of the
Type IIn supernova (SN IIn) 2010jl. The SN is undetected by {\it
  SOFIA}, but the upper limits obtained, combined with new and
archival detections from {\it Spitzer} at 3.6 \& 4.5 $\mu$m allow us
to characterize the composition of the dust present. Dust in other
Type IIn SNe has been shown in previous works to reside in a
circumstellar shell of material ejected by the progenitor system in
the few millenia prior to explosion. Our model fits show that the dust
in the system shows no evidence for the strong, ubiquitous 9.7 $\mu$m
feature from silicate dust, suggesting the presence of carbonaceous
grains. The observations are best fit with 0.01-0.05 $\msun$ of
carbonaceous dust radiating at a temperature of $\sim 550-620$ K. The
dust composition may reveal clues concerning the nature of the
progenitor system, which remains ambiguous for this subclass.  Most of
the single star progenitor systems proposed for SNe IIn, such as
luminous blue variables, red supergiants, yellow hypergiants, and B[e]
stars, all clearly show silicate dust in their pre-SN
outflows. However, this post-SN result is consistent with the small
sample of SNe IIn with mid-IR observations, none of which show signs
of emission from silicate dust in their IR spectra.


\end{abstract}

\section{Introduction}
\label{intro}

Type IIn supernovae (SNe IIn) are named for the narrow hydrogen
emission lines produced in a dense, slowly-moving, pre-existing
circumstellar medium (CSM).  Typically attributed to the progenitor
winds, the CSM has densities and velocities that suggest extremely
high mass-loss rates, in some cases up to $\sim 0.1 \msun$ yr$^{-1}$
(\citealt{fox11}, hereafter F11; \citealt{kiewe12}). The progenitors
of SNe IIn are unknown. The implied mass-loss rates are often compared
to the episodic dense winds observed in some massive stars, such as
Luminous Blue Variables \citep[LBVs; e.g.,][and references
  therein]{smith11}.

SNe IIn account for more than half of all known SNe with late-time ($>
100$ days) infrared (IR) emission (F11), implying the presence of warm
dust. F11 showed, via a warm mission (3.6 and 4.5 $\mu$m only) {\it
  Spitzer} survey of 10 SNe IIn, that the observed emission is
consistent with pre-existing CSM dust, heated by the optical emission
generated by the forward shock interaction with the dense CSM. This is
in contrast to most core-collapse (CC) SNe with IR emission, where
observations are consistent with a small amount of newly-formed ejecta
dust \citep[e.g.,][]{kotak05,sugerman06,williams08}.  Since the dust
is produced in the outflow of the progenitor system, characterizing
this dust may offer a clue about the pre-SN system. However, as F11
showed, {\it Spitzer} observations at 3.6 and 4.5 $\mu$m are
sufficient only to establish the presence of a warm dust component,
and not to determine the composition of the dust.

It is well-known that massive stars make dust in their stellar winds,
but what {\it type} of dust depends on the star's mass and
evolutionary state. In general, stars make either silicate or carbon
dust in their outflows, where the determining factor is the C/O ratio
in the stellar atmosphere. The presence or absence of a broad, strong
emission feature at 9.7 $\mu$m that only arises from silicate dust
provides a method by which to distinguish the two grain populations,
provided images or spectra exist in the 8-13 $\mu$m range.

\subsection{Potential SNe IIn Progenitors and their Dust Signatures}
\label{potential}

The LBV scenario is a promising candidate for the progenitor systems
of SNe IIn. \citet{galyam09} identify a hypergiant progenitor
undergoing LBV-like mass loss in pre-explosion images of SN
2005gl. However, it has several complications.  Most LBVs achieve
observed rates of only $\mdot < 10^{-4}$~\ml~in their S Doradus state
\citep{humphreys94}.  Standard radiation and line-driven wind models
similarly have difficulty reproducing such substantial rates
\citep[e.g.,][]{smith06}. Classical stellar evolution models suggest
these massive stars should evolve into the Wolf-Rayet (WR) phase
before exploding \citep{heger03}.  More recently, proposed solutions
include swept up Wolf-Rayet winds \citep{dwarkadas11}, binary
accretion \citep{kashi10}, and gravity-wave driven winds
\citep{quataert12}, but these all have limitations of their own.

While often variable and unpredictable in brightness, LBVs are
typically very luminous in the IR due to the significant amounts of
dust formed in their outflows. Though hardly a homogeneous bunch, most
LBVs are consistent with some sort of silicate dust in their
spectra. Some, such as HR Car \citep{umana09}, show strong features
from amorphous silicate grains, while others, like the Large
Magellanic Cloud (LMC) LBV R71, exhibit features of crystalline
silicates \citep{morris08}. Additionally, a few LBVs like HDE 316285
and HD 168625 \citep{morris08,umana09}, show emission features from
polycyclic aromatic hydrocarbons (PAHs), though whether these PAHs are
formed in the stellar outflow or are pre-existing in the interstellar
medium (ISM) is unknown. In $\eta$ Carinae, \citet{smith10} report
that the IR spectra may be best fit with corundum dust
(Al$_{2}$O$_{3}$), which also shows a strong $\sim 10$ $\mu$m
feature. We are unable to identify any examples in the literature of
IR spectra from CSM surrounding an LBV that unambiguously shows signs
of graphite or amorphous carbon dust.

Red supergiants (RSGs), typically very bright in the IR, also show
signs of silicate dust in the spectra of their CSM. In a sample of 19
RSGs in the LMC, \citet{woods11} report that amongst the 14 that show
emission from warm dust, most show the 9.7 $\mu$m silicate feature, or
other features of non-carbonaceous dust. They note that a few show
weak PAH emission features, but that at least some of these may arise
from the ISM.

Other possibilities include any massive star that loses a significant
amount of mass to stellar winds after evolving off the
main-sequence. Yellow hypergiants are believed to be intermediate
stages of stellar evolution with strong winds, between the RSG and LBV
phases \citep{lagadec11}. Again, though, the dusty shells seen around
some of these stars, like IRC+10420, show strong silicate features in
their spectra \citep{humphreys97}. B[e] supergiants, massive ($\sim
15-70\ \msun$), luminous, poorly understood early-type stars with slow
stellar winds that form dusty circumstellar material may also be
precursors to LBVs and/or WR stars. A {\it Spitzer} spectroscopic
study of B[e] supergiants in the LMC shows that the dusty disks around
these stars contain strong features of both crystalline and amorphous
silicates, along with PAH features, but no carbonaceous dust
\citep{kastner06}.

An intriguing potential progenitor is suggested by \citet{wesson10}
and \citet{kochanek11} based on archival {\it Spitzer} observations of
the possible Type IIn SN 2008S. They rule out silicate grains and
state that this absence of silicates may be evidence against a
supergiant progenitor, favoring instead an ``extreme'' asymptotic
giant branch (AGB) star. However, this result should be interpreted
with caution, as other authors report that SN 2008S is in fact a
supernova impostor, and was more likely an LBV eruption
\citep{smith09}. SN impostors and/or LBV eruptions may be relevant if,
ultimately, they explode as SNe IIn \citep[see][]{mauerhan13}.

Some authors have proposed a binary scenario for the progenitors of
SNe IIn. Among binary systems, observations of carbon-rich stars AGB
stars are consistent with mass transfer from an evolved binary
companion \citep{aoki02}. In order to explain the apparent close
timing between the mass loss and SN events, \citet{chevalier12}
explored the possibility of mass loss driven by a common envelope
involving a compact object and a massive star, finding that at least
some SNe IIn could be explained this way.

$\eta$ Car itself is a binary system. \citet{iping05} report the
likely detection of the companion star, $\eta$ Car B, and note that
while a detailed stellar classification for this secondary is
inconclusive, $\eta$ Car B could potentially be a Wolf-Rayet
star. Neither the formation site nor the composition of the dust in
$\eta$ Car is well-understood \citep{smith10}. Even the origin of the
CSM is uncertain, as winds from either or both of the two stars could
contribute.

\subsection{SNe IIn in the IR}

Only a few SNe IIn have been observed and detected at mid-IR ($>$ 4.5
$\mu$m) wavelengths. \citet{fox10} show 5-14 $\mu$m {\it Spitzer} IRAC
and IRS observations of SN 2005ip, whose spectrum is clearly fit with
a carbonaceous grain model with no contribution from silicate dust
required. \citet{vandyk13} detected SN 1995N in archival data at 12
$\mu$m with the {\it Wide-Field Infrared Survey Explorer, (WISE)} and
24 $\mu$m with {\it Spitzer}, and showed that both silicate and
graphite dust models have problems fitting the IR spectral energy
distribution (SED). Silicate grains were also ruled out for SN 2008S,
which was detected at 24 $\mu$m by {\it Spitzer}, though its status as
a supernova is questionable (see Section~\ref{potential}). SN 2006jd
was seen at 12 $\mu$m by the {\it Wide-Field Infrared Survey Explorer
  (WISE)} \citep{stritzinger12} without the silicate feature.

SN 2010jl was observed from the ground in the optical/near-IR by
\citet{maeda13}. Their data is sensitive to the hottest component of
the dust, and they find dust temperatures of $> 1000$ K. While they
find equally good fits to this portion of the spectrum with
carbonaceous and silicate grains, the temperatures required for the
silicate grains are higher than the condensation temperature for those
grains, and thus they favor carbonaceous grains. In this {\it Letter},
we report direct observations of the all-important mid-IR band using
the {\it Stratospheric Observatory for Far-Infrared Astronomy
  (SOFIA)}, where we can detect the presence or lack of a silicate
bump. We do not detect SN 2010jl, but the upper limits we derive rule
out the presence of a silicate feature, and are well-fit by models of
carbon dust, confirming the conclusions of \citet{maeda13}.

\section{Observations}

SN 2010jl was discovered in UGC 5189A (distance 49 Mpc) on 2010
Nov. 3.5 by \citet{newton10}, at a position of $\alpha = 09^{\rm h}
42^{\rm m} 53.^{\rm s}337$, $\delta = +09^{\circ}29'42''.13$
(J2000.0). It was identified as a Type IIn by \citet{benetti10} and
\citet{yamanaka10}. Multi-wavelength observations all point to a
strongly interacting SN with a dense CSM
\citep{fransson14,ofek14}. Pre-SN {\it Hubble Space Telescope} images
show a luminous, blue point source at the position of the SN
\citep{smith11}, but the data are insufficient to determine whether
this source is a young cluster, a single luminous star, or a star
caught during outburst. SN 2010jl shows significant evidence for
late-time IR emission. Some interpret this as evidence for newly
formed dust \citep[e.g.][]{smith12,maeda13,gall14,jencson15}, while
others suggest the presence of a pre-existing, unshocked dust shell
\citep[e.g.,][]{andrews11,fox13,fransson14,borish15}.

\begin{figure}
\includegraphics[width=15cm]{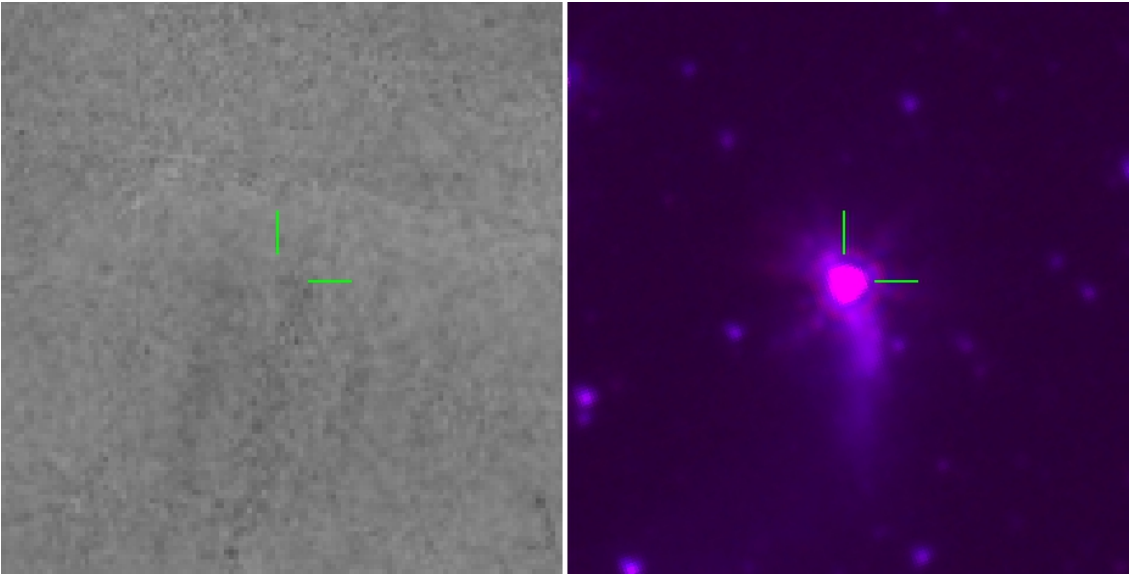}
\caption{{\it Left}: {\it SOFIA} 11.1 $\mu$m image of the field of SN
  2010jl. Neither the SN nor the host galaxy are detected. {\it
    Right}: Spitzer 3.6 (blue) \& 4.5 (red) $\mu$m image of the same
  region, clearly showing the SN, indicated by the tick marks, and
  host galaxy UGC 5189A.
\label{images}
}
\end{figure}

As part of an ongoing monitoring project of several Type IIn SNe at
late times (PI: O. Fox), SN 2010jl has been observed in the mid-IR
roughly 10 times with the {\it Spitzer Space Telescope}, all at the
near-IR wavelengths of 3.6 and 4.5 $\mu$m (the SN occurred after the
end of {\it Spitzer's} cryogenic mission). The SN is clearly detected
in the {\it Spitzer} images, as we show in Figure~\ref{images}, and
has a flux of several mJy at both wavelengths. We also observed the SN
with the FORCAST instrument on {\it SOFIA} using the 11.1 $\mu$m
filter on 2014 May 5-6 at an altitude of 38000 feet, with a total
integration time of 6400 s. The images are taken in chop/nod mode with
a 60$''$ throw, and several hundred 30 s frames were coadded for a
final data mosaic. The final mosaic has pixels 0.75$''$ on a side with
a resolution of approximately 2.7$''$.

Neither SN 2010jl nor the host galaxy were detected in our mosaicked
image. The lack of detection of the host galaxy makes setting an upper
limit on any point source easier, since we are background-limited and
not confusion-limited. To obtain a 1$\sigma$ upper limit, we computed
the standard deviation of the pixels in a $20'' \times 20''$ box
centered on the source coordinates and converted this via the CALFCTR
header parameter to get an RMS value in Jy/pix. We then multiplied
this by the square root of the extraction area (4 pixel radius, 50
total pixels) centered on SN 2010jl. The upper limit obtained in this
method is 4.2 mJy, in good agreement with the 4.7 mJy value that we
extrapolate from the {\it SOFIA} Observer's Handbook (Section 7.1.4)
based on our exposure time.

Since SNe are time-variable objects, we must compare our {\it SOFIA}
observations with those most contemporaneous from {\it Spitzer}. We
extract fluxes for four of the {\it Spitzer} observations, taken on
2013 Jan 30, 2013 Jun 29, 2013 Jul 3, and 2014 Jul 9. We find no clear
trend of systematic brightening or fading, so we average these four
fluxes for each wavelength and use this average flux for our modeling,
below. The average fluxes are 5.0 (1.4) and 5.69 (1.3) mJy for 3.6 and
4.5 $\mu$m, respectively. The values in parentheses represent the
standard deviation of the four measurements, which are significantly
larger than {\it Spitzer's} photometric uncertainties, and thus a very
conservative estimate of the uncertainties.

\section{Dust Modeling with the New {\it SOFIA} Upper Limit}

F11 modeled the {\it Spitzer} 3.6 \& 4.5 $\mu$m emission from multiple
SNe IIn, considering a variety of origins and heating mechanisms for
the dust. They rule out newly formed ejecta dust, as well as
collisional heating of CSM dust by hot electrons behind the forward
shock. They find that the most likely physical scenario is that a
pre-existing dust shell is heated by UV and optical radiation from
circumstellar interaction of the forward shock with dense
material. The two short wavelength data points are insufficient to
determine the type of dust present, and F11 considered both silicate
and graphite grains. With the addition of the {\it SOFIA} upper limit
at 11.1 $\mu$m, we can compare models of silicate and graphite dust
over a much broader wavelength range that includes the 9.7 $\mu$m
silicate feature.

The spectrum of a warm dust grain is simply the wavelength-dependent
absorption cross-section of the grain multiplied by the Planck
blackbody function. We consider several different types of grains. In
addition to graphite and ``astronomical'' silicate grains, with
absorption cross-sections derived from bulk optical constants given in
\citet{draine84}, we also consider amorphous carbon grains
\citep{rouleau91}, as well as other silicate grains, such as glassy
and amorphous forms of both enstatite and forsterite
\citep{jager03,dorschner95}. For all grains, we assume a single grain
size of 0.1 $\mu$m. This model is simplistic, but a more detailed
model cannot be constrained by only three data points, and the overall
shape of the spectrum is not significantly changed by the addition of
more grain sizes. More importantly, \citet{temim13} have recently
shown that more physically realistic dust models, which take into
account a range of grain-size distributions, will lower the amount of
mass inferred. Thus, our models represent an {\it upper-limit} to the
amount of dust formed in the pre-existing CSM shell.

\begin{figure}
\includegraphics[width=16cm]{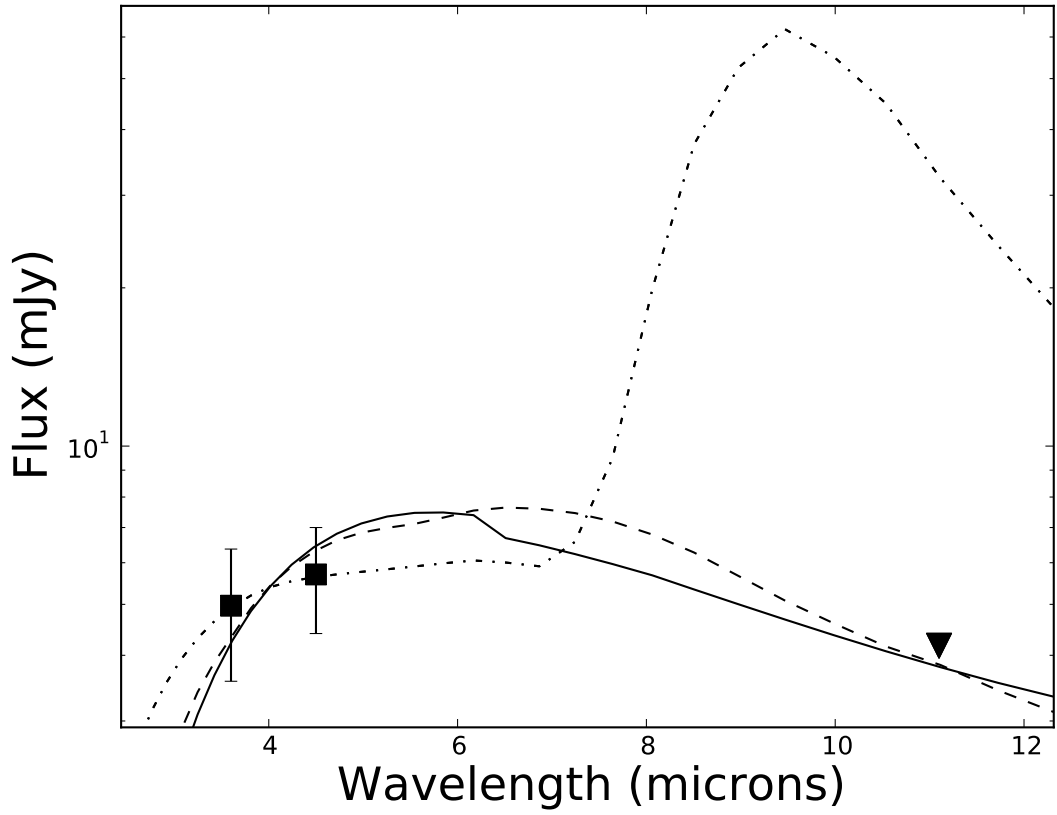}
\caption{Model fits to the data for graphite (solid black line),
  amorphous carbon (dashed line), and silicate dust (dot-dashed
  line). {\it Spitzer} 3.6 and 4.5 $\mu$m fluxes are shown as squares,
  and the 1$\sigma$ {\it SOFIA} upper limit is marked with a
  triangle. Graphite and amorphous carbon dust models are fit to all
  three data points (assuming {\it SOFIA} upper limit is a detection),
  while the silicate dust model is fit only to the two {\it Spitzer}
  points. The best-fit silicate model overpredicts the 11.1 $\mu$m
  emission by roughly an order of magnitude.
\label{modelfits}
}
\end{figure}

We use a least-squares algorithm to fit the data for each grain
composition. Since the flux values are of the same order of magnitude,
we fit in linear space, and use only two free parameters: dust
temperature and total mass. We obtain good fits to the SED with models
of both graphite (T = 553 K) and amorphous carbon (T = 616 K) grains,
as we show in Figure~\ref{modelfits}. While we cannot further
distinguish between these two models, the difference is not
particularly relevant to the goals of this paper. The only significant
difference between the graphite and amorphous carbon models is the
amount of dust needed to fit the spectrum. Since amorphous carbon has
a higher emissivity than graphite over the range of $\sim 1-20\ \mu$m,
it requires less dust to achieve the same luminosity. Our graphite
model requires 0.057 $\msun$ of dust, while only 0.012 $\msun$ of
amorphous carbon is necessary (assuming 49 Mpc distance).

Most importantly, we find that no silicate grain model can adequately
reproduce the data for all three points. We show the best-fit model
for ``astronomical silicates'' in Figure~\ref{modelfits}, as fit to
the two {\it Spitzer} points alone. This model requires a dust
temperature of 863 K with a radiating dust mass of 0.024 $\msun$. The
11.1 $\mu$m model prediction from this model is roughly an order of
magnitude higher than the upper limit of the flux density from the
{\it SOFIA} data. Even fitting a silicate model to the extreme
uncertainty values of the {\it Spitzer} fluxes yields completely
nonphysical results. Other silicate models, such as ``glassy''
silicates, lead to even poorer fits. While we cannot rule out a small
contribution from silicate grains (having only three data points makes
this difficult to constrain), silicate dust does not dominate the SED,
and it is not required at all to obtain a good fit to the data. If the
dust we see comes from a pre-existing shell of material, then this
conclusion is at odds with the pre-SN dust observed in most proposed
progenitor systems for SNe IIn.

WR stars are one type of massive star that is known to produce
significant amounts of carbon dust \citep{woods11}. However, since WR
stars have blown off their hydrogen (and in some cases, helium)
envelopes, it is generally accepted that WR stars would explode as
Type Ib/c SNe. An interesting suggestion is made by
\citet{dwarkadas11}, who argues that at least some SNe IIn may explode
after brief ($< 10^{4}$ yr) WR phases for their progenitors. This is
quantitatively consistent with the model of F11, who report that the
progenitor system of SN 2006jd experienced a significant period of
mass loss a few hundred years prior to explosion. Work by
\citet{anderson12} found, by investigating the correlation of SNe IIn
with the recent star formation history of the explosion site, that SNe
IIn trace recent, but not ongoing star formation, implying that most
of these SNe do not arise from the most massive stars.

There are other possibilites as well. The dust observed may be from
newly-formed ejecta dust, and thus not indicative of the pre-SN
outflow. A high optical depth or significantly larger grains than are
found in the ISM might suppress the silicate feature
\citep{lucy91}. While we cannot say for certain that the lack of a
silicate feature definitively means a carbon rich environment, the
results here and in other SNe IIn are interesting and merit further
study.

\section{Summary}

The Type IIn SN 2010jl, easily seen in {\it Spitzer} 3.6 and 4.5
$\mu$m, is undetected in 11.1 $\mu$m images from {\it SOFIA}. With the
flux at the two {\it Spitzer} wavelengths and an upper limit at 11.1
$\mu$m, we can characterize the dust composition. Only carbonaceous
grains (either graphite or amorphous carbon) can fit the spectrum;
silicate grain models overpredict the 11.1 $\mu$m upper limit by an
order of magnitude or more. Our model fits suggest on the order of
0.01-0.05 $\msun$ of dust radiating at a temperature of $\sim 550-620$
K. SN 2010jl joins a growing list of SNe IIn whose dust signatures do
not show the 9.7 $\mu$m silicate feature. As F11 argue, the dust in
these SNe is contained in a pre-existing circumstellar shell, blown
off by the progenitor system before the explosion. Thus, the dust that
we observe can reveal something about the unknown nature of the
progenitor.

Most of the potential progenitor classes for SNe IIn are towards the
massive end of the stellar spectrum: LBVs, RSGs, yellow hypergiants,
and B[e] stars, all of which eject significant quantities of
material. However, they are all observed to produce silicate dust
during their pre-SN lifetimes, not carbonaceous. It is possible some
stars may have a mixed chemistry in their outflows, producing some
amount of both silicate and carbon grains. This has been observed in
some AGB stars, and \citet{williams12} inferred the presence of dust
grains of both types in {\it Spitzer} observations of Kepler's
supernova remnant. Although Kepler is the remnant of a Type Ia, it is
known to be interacting with a dense CSM from the progenitor system,
and could be a low mass analog of SNe IIn. An intriguing possibility
is that of a binary system where one of the stars is a carbon-rich WR
star.

While we advise caution against drawing broad conclusions about the
nature of SNe IIn from an indirect method and such a small sample
size, we urge further study of these systems from the standpoint of
both stellar evolution and dust nucleation in the atmospheres and
outflows of massive stars. The {\it James Webb Space Telescope} will
enable spectroscopic detection of many SNe IIn in the mid-IR, allowing
detailed studies of individual SN spectra, as well as the ability to
do statistical studies on large samples of SNe IIn in the mid-IR.

{\it Facilities:} \facility{SOFIA}, \facility{Spitzer}

\acknowledgements

We thank Nathan Smith for useful discussions, and the anonymous
referee for manyhelpful suggestions. BJW thanks {\it SOFIA} for the
opportunity to fly on the aircraft during the observing run.

\newpage
\clearpage

\begin{thebibliography}{}

\bibitem[Anderson et al.(2012)]{anderson12}
Anderson, J.P., Habergham, S.M., James, P.A., \& Hamuy, M. 2012, MNRAS, 424, 1372

\bibitem[Andrews et al.(2011)]{andrews11}
Andrews, J.E., et al. 2011, AJ, 142, 45

\bibitem[Aoki et al.(2002)]{aoki02}
Aoki, W., Norris, J.E., Ryan, S.G., Beers, T.C., \& Ando, H. 2002, ApJ, 567, 1166

\bibitem[Benetti et al.(2010)]{benetti10}
Benetti, S., et al. 2010, CBET, 2536, 1

\bibitem[Borish et al.(2015)]{borish15}
Borish, H.J., Huang, C., Chevalier, R.A., Breslauer, B.M., Kingery, A.M., Privon, G.C. 2015, ApJ, 801, 7

\bibitem[Chevalier(2012)]{chevalier12}
Chevalier, R.A. 2012, ApJ, 752, 2

\bibitem[Dorschner et al.(1995)]{dorschner95}
Dorschner, J., Begemann, B., Henning, T., J$\ddot{\rm a}$ger, C., \& Mutschke, H., 1995, A\&A, 300, 503

\bibitem[Draine \& Lee(1984)]{draine84}
Draine, B.T. \& Lee, H.M. 1984, ApJ, 285, 89

\bibitem[Dwarkadas(2011)]{dwarkadas11}
Dwarkadas, V.V. 2011, MNRAS, 412, 1639

\bibitem[Fox et al.(2010)]{fox10}
Fox, O.D., et al. 2010, ApJ, 725, 1768

\bibitem[Fox et al.(2011)]{fox11}
Fox, O.D., et al. 2011, ApJ, 741, 7

\bibitem[Fox et al.(2013)]{fox13}
Fox, O.D., et al. 2013, AJ, 146, 2

\bibitem[Fransson et al.(2014)]{fransson14}
Fransson, C., et al. 2014, ApJ, 797, 118

\bibitem[Gall et al.(2014)]{gall14}
Gall, C., et al. 2014, Nature, 511, 326

\bibitem[Gal-Yam \& Leonard(2009)]{galyam09}
Gal-Yam, A., \& Leonard, D.C. 2009, Nature, 458, 865

\bibitem[Heger et al.(2003)]{heger03}
Heger, A., Fryer, C.L., Woosley, S.E., Langer, N., \& Hartmann, D.H. 2003, ApJ, 591, 288

\bibitem[Humphreys \& Davidson(1994)]{humphreys94}
Humphreys, R.M., \& Davidson, K. 1994, PASP, 106, 1025

\bibitem[Humphreys et al.(1997)]{humphreys97}
Humphreys, R.M., et al. 1997, AJ, 114, 2778

\bibitem[Iping et al.(2005)]{iping05}
Iping, R.C., Sonneborn, G., Gull, T.R., Massa, D.L., \& Hillier, D.J. 2005, ApJ, 633, 37

\bibitem[J$\ddot{\rm a}$ger et al.(2003)]{jager03}
J$\ddot{\rm a}$ger, C., Dorschner, J., Mutschke, H., Posch, T., \& Henning, T., 2003, A\&A, 408, 193

\bibitem[Jencson et al.(2015)]{jencson15}
Jencson, J.E., Prieto, J.L, Kochanek, C.S., Shappee, B.J., Stanek, K.Z., \& Pogge, R.W. 2015, arxiv::1505.01186

\bibitem[Kashi(2010)]{kashi10}
Kashi, A. 2010, AIP Conf. Proc. 1314, 55

\bibitem[Kastner et al.(2006)]{kastner06}
Kastner, J.H., Buchanan, C.L., Sargent, B., \& Forrest, W.J. 2006, ApJ, 638, 29

\bibitem[Kiewe et al.(2012)]{kiewe12}
Kiewe, M., et al. 2012, ApJ, 744, 10

\bibitem[Kochanek(2011)]{kochanek11}
Kochanek, C.S. 2011, ApJ, 741, 37

\bibitem[Kotak et al.(2005)]{kotak05}
Kotak, R., Meikle, P., van Dyk, S.D., H$\ddot{\rm o}$flich, P.A., \& Mattila, S. 2005, ApJ, 628, 123

\bibitem[Lagadec et al.(2011)]{lagadec11}
Lagadec, E., et al. 2011, A\&A, 534, 10

\bibitem[Lucy et al.(1991)]{lucy91} Lucy, L.B., Danziger, I.J.,
  Gouiffes, C., \& Bouchet, P. 1991, in Supernovae. The Tenth Santa
  Cruz Workshop in Astronomy and Astrophysics, held July 9-21, 1989
  Lick Observatory. Editor, S.E. Woosley; Publisher, Springer-Verlag,
  New York, 1991. LC \# QB856 .S26 1989. ISBN \# 0387970711. P. 82, 1991

\bibitem[Maeda et al.(2013)]{maeda13}
Maeda, K., et al. 2013, ApJ, 776, 5

\bibitem[Mauerhan et al.(2013)]{mauerhan13}
Mauerhan, J.C. et al. 2013, MNRAS, 430, 1801

\bibitem[Morris et al.(2008)]{morris08}
Morris, P., et al. 2008, in ``Massive Stars as Cosmic Engines,'' ed. F. Bresolin, P.A. Crowther, \& J. Puls, Proceedings IAU Symposium No. 250

\bibitem[Newton \& Puckett(2010)]{newton10}
Newton, J., \& Puckett, T., CBET, 2532, 1

\bibitem[Ofek et al.(2010)]{ofek14}
Ofek, E.O., et al. 2014, ApJ, 781, 42

\bibitem[Quataert \& Shiode(2012)]{quataert12}
Quataert, E., \& Shiode, J. 2012, MNRAS, 423, 92

\bibitem[Rouleau \& Martin(1991)]{rouleau91}
Rouleau, F. \& Martin, P.G. 1991, ApJ, 377, 526

\bibitem[Smith \& Owocki(2006)]{smith06}
Smith, N., \& Owocki, S.P. 2006, ApJ, 645, 45

\bibitem[Smith et al.(2009)]{smith09}
Smith, N. et al. 2009, ApJ, 697, 49

\bibitem[Smith(2010)]{smith10}
Smith, N. 2010, MNRAS, 402, 145

\bibitem[Smith et al.(2011)]{smith11}
Smith, N., Li, W., Silverman, J.M., Ganeshalingam, M., Filippenko, A.V. 2011, MNRAS, 415, 773

\bibitem[Smith et al.(2012)]{smith12}
Smith, N. 2012, AJ, 143, 17

\bibitem[Stritzinger et al.(2012)]{stritzinger12}
Stritzinger, M., et al. 2012, ApJ, 756, 173

\bibitem[Sugerman et al.(2006)]{sugerman06}
Sugerman, B.E.K., et al. 2006, Science, 313, 196

\bibitem[Temim \& Dwek(2013)]{temim13}
Temim, T. \& Dwek, E. 2013, ApJ, 774, 8

\bibitem[Umana et al.(2009)]{umana09}
Umana, G., Buemi, C.S., Trigilio, C., Hora, J.L., Fazio, G.G., \& Leto, P. 2009, ApJ, 694, 697

\bibitem[van Dyk(2013)]{vandyk13}
van Dyk, S. 2013, AJ, 145, 118

\bibitem[Wesson et al.(2010)]{wesson10}
Wesson, R., et al. 2010, MNRAS, 403, 474

\bibitem[Williams et al.(2008)]{williams08}
Williams, B.J., et al. 2008, ApJ, 687, 1054

\bibitem[Williams et al.(2012)]{williams12}
Williams, B.J., et al. 2012, ApJ, 755, 3

\bibitem[Woods et al.(2011)]{woods11}
Woods, P.M. et al. 2011, MNRAS, 411, 1597

\bibitem[Yamanaka et al.(2010)]{yamanaka10}
Yamanaka, M., Okushima, T., Arai, A., Sasada, M., Sato, H., CBET, 2539, 1

\end{thebibliography}
\end{document}